\documentclass[aps,pre,longbibliography,reprint]{revtex4-2}

\usepackage{graphicx}
\usepackage{dcolumn}
\usepackage[dvipsnames]{xcolor}

\usepackage{xcolor}
\usepackage{amsmath}
\usepackage{changes}
\setdeletedmarkup{\color{red}{\sout{#1}}}
\usepackage{bm}
\DeclareUnicodeCharacter{0301}{\'o}

\begin{document}

\title{Edge slip stabilizes confined active vortices by suppressing localized instabilities}

 \author{Zhihan Ye}
 \author{Tianyu Ren} 
 \author{Hao Luo}
 
 \author{Yanan Liu}
 \email{yanan.liu@nwu.edu.cn}

 \author{Guangyin Jing}
 \email{jing@nwu.edu.cn}
 \affiliation{School of Physics, Northwest University}
 \affiliation{Peng Huanwu Center for Fundamental Theory, Shaanxi Key Laboratory for Theoretical Physics Frontiers, Northwest University, 710127, Xi’an, China}

\date{\today}

\begin{abstract}
Confined active systems can sustain persistent vortical flows whose stability is strongly influenced by boundary conditions.
At the individual level, active units generate internal stresses that drive spontaneous flows, which in turn advect and reorient the particles. 
This nonlinear coupling between active flow and orientational order is significantly mediated by the system's boundaries, where the specific slip condition governs how these internal stresses generate active flow then rearrange the local orientations. 
However, a quantitative understanding of how boundary slip dictates their dynamical stability remains lacking. 
Here, we study how the slip boundary condition controls the stability of a steady vortex state in a circularly confined active nematic system. 
Using a continuum model in a flow-dominated regime, we perform a linear stability analysis and derive an explicit criterion incorporating the slip velocity and flow-alignment coupling. 
We find that increasing slip velocity suppresses localized linear instabilities, thereby promoting the persistence of the steady vortex state. 
This reveals a relaxing the boundary friction actually stabilizes the macroscopic coherent structure by depressing flow induced reorientation that typically destroys single-vortex states. 
Our findings establish boundary slip as a nontrivial hydrodynamic control parameter for engineering stable active flows.
\end{abstract}

\maketitle

\clearpage
\section{\label{sec:Intro}Introduction}
Active systems consist of energy-consuming, self-driven units that exhibit persistent motion~\cite{introduce_active_matter0}. 
They are found across scales, from bacterial suspensions and self-propelled colloids to animal flocking in nature and robotic swarms in artificial realizations~\cite{introduce_active_matter}.
These systems exhibit a rich variety of emergent collective phenomena, including spontaneous flows, vortices, and topological defects~\cite{introduce_active_turbulence,introduce_active_turbulence2,introduce_active_turbulence3,Long_wave_instability,Long_wave_instability2,Long_wave_instability3}. 
One category of active systems, driven by active stress to generate large-scale flow and composed of apolar components, is called active nematic systems, such as cytoskeletal filaments and microtubules driven by protein motors~\cite{Spotounes_instability}. 
These active units reorient themselves by the self-generated flows, and therefore regulate the active flow with the self-adjusting orientation order, resulting in a nonlinear coupling between flow and orientation~\cite{introduce_active_nematics,introduce_active_nematics2,introduce_active_nematics3,introduce_active_nematics4}.
Typically, these coordinated motions generate coherent flow structures, leading to a defect-mediated turbulent state in an unbounded system ~\cite{introduce_nematic_turbulence1,introduce_nematic_turbulence2,introduce_nematic_turbulence3}.
In experiments, dense elongated bacteria or microtubules with nematic alignment have shown the rapid formation and annihilation of defects, analogous to liquid crystals~\cite{Introduce_bacterial_turbulence}. 
Vortical flow accompanies these defects, while also perturbing and breaking the local orientation order.

These intricate coupled nonlinear effects perpetually disrupt coherent structures most prominently in unbounded systems.
However, biological systems in nature are invariably bounded or even capable of spontaneously generating boundaries~\cite{theory_of_cell_defect,onset_of_collective_motion,MIPS}.
More remarkably, active matter continuously interacts with boundaries, giving rise to stabilized ordered structures that are difficult to maintain in boundary-free systems.
One approach towards ordered active flows involves modulating active particles within quasi-two-dimensional confinement through either dissipative boundaries imposed by upper and lower walls or pre-embedded patterns to mechanically guide active structures~\cite{mechanisim_of_artificial_nematic_pattern,artificial_nematic_pattern,artificial_nematic_pattern_of_cell,artificial_nematic_pattern2,mehcanics_of_artificial_nematic_pattern2,mechanical_regulation,Mechanical_Regulation1,Mechanical_Regulation2}.
Such methods can be applied to control active transport and enhance wound healing. 
Another more ubiquitous natural scenario involves systems constrained not only in height but also possessing edges—analogous to bacterial suspension inhabiting soil or droplets. 
Induced by the lateral confinement, chiral symmetry breaking spontaneously occurs in such systems, leading to the formation of a single integer defect accompanied by circulating flows~\cite{Bacteria_vortex_RE_exp,Bacteria_vortex_RE_sim,edge_control_vortex,vortex_in_edge_confined_space3}.

Assessing how edge properties influence the dynamics of active matter has attracted considerable research interest.
Recent studies have revealed that edge geometry can induce edge-protected flows, and that soft boundaries with surface tension, such as those arising in cellular tissues or developing embryos, compete with active stresses to yield distinct morphological patterns~\cite{edge_flow,soft_boundary,cell_vortex_in_confined_space,cell_vor_in_confined_space_2}.
An important question arises concerning the critical system size for  a single active vortex.
David Santillan et al. addressed this problem in an early study, by employing kinetic theory to quantitatively link self-organized steady states to activity and confinement effects~\cite{David_vortex_theory}.
Recently, it was reported that active particles in a confined well can  transition from a single vortex to chaotic patterns at a critical system size ~\cite{vortex_in_edge_confined_space2,vortex_instability}. 
One of these works has reported that by varying the system size, the nucleation of a pair of positive and negative defects near the edge triggers the breakdown of the single vortex~\cite{boundary_friction}.
In addition, recent work by B. Perez-Estay et al. reported a giant single bacterial vortex confined within two glass plates while surrounded by gas-liquid side boundary, and confirmed that this single vortex could sustain for a longer time than which surrounded by solid-liquid boundary~\cite{scale_free}. 
This reminds us that the slip boundary plays a crucial role in the stability of the active flow and vortex structures under confinement.
However, there is still a lack of quantitative discussion on how the slip boundary condition governs the stability and the steady state of the active vortex under confinement.

\begin{figure}
    \centering
    \includegraphics[width=0.8\linewidth]{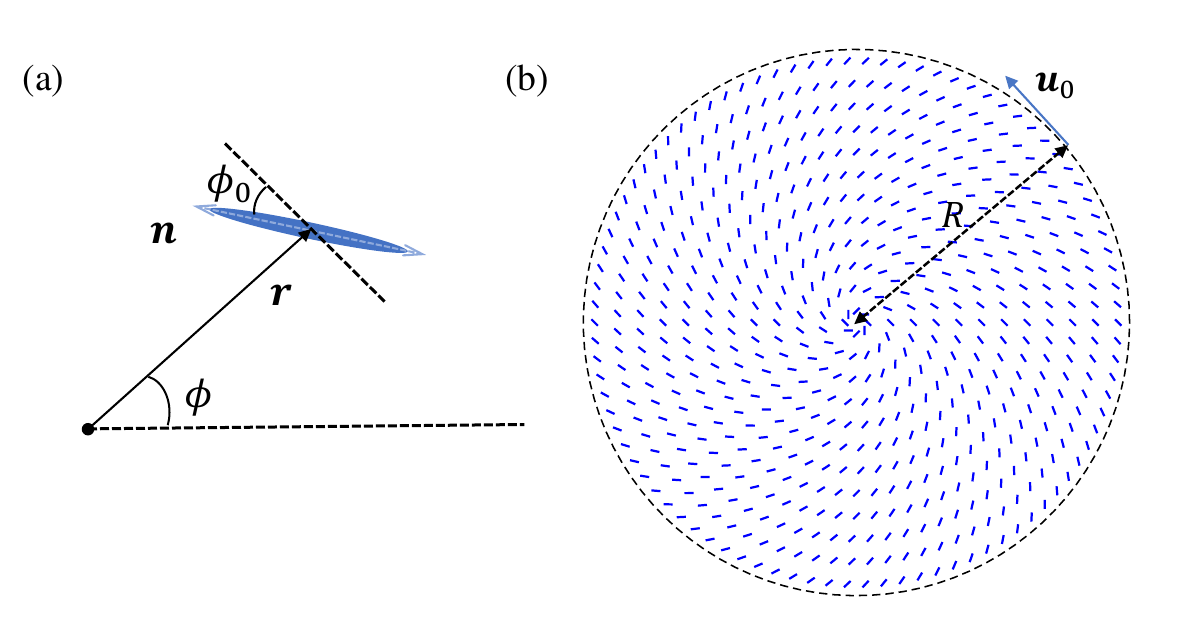}
    \caption{\textbf{Sketch of problem setting.}(a) A nematic particle under polar coordination, with position $\mathbf{r}=(r,\phi)$,  orientation direction $\mathbf{n}$, and the angle $\phi_{0}$ between tangential direction and orientation direction. 
    (b) A single vortex with spiral configuration with radius $R$, and the slip velocity \added{$u_{0}$} at the edge.}
    \label{fig:defect sketch}
\end{figure}

In this paper, we start from an integer nematic defect, i.e. a single vortex, and employ the active liquid crystal model to investigate how the slip boundary condition affects the linear stability of a pre-existing vortex.
First, by drawing an analogy with stability of a non-viscous fluid, the criterion for a stable vortex is shown to reduce to is recovered to the classical Rayleigh criterion for rotating flow. 
Then we conduct a linear stability analysis on the vortex, demonstrating that increasing slippery boundary is beneficial for the stabilization of an active single vortex, the nematic defect with integer topological charge.
Furthermore, we examine the quantitative dependence on the flow alignment of active particle, and determine a specific flow-orientation coupling strength, which the classical vortex stability criterion holds.
The stability control here suggests a dynamical treatment for coherent structures consisting of living units in population, including cell clusters in confined environments, and could inform the design of stably tunable active vortices in experimental systems.

\section{\label{sec:Model}Continuum model}\label{II}

For these stress-driven nematic systems, the active liquid crystal theory accurately characterizes the large-scale dynamics~\cite{Half_Skyrmion,introduce_active_nematics,Free_Energy_of_Q,ALCmodel}, through equations of motion for the velocity and nematic fields
The nematic order parameter is given by $\mathbf{Q}=S \left(\mathbf{nn}-\frac{\mathbf{I}}{2}\right)$, with unit direction $\mathbf{n}$. 
The flow field $\mathbf{u}(\mathbf{r})$, with the incompressible assumption $\nabla \cdot \mathbf{u}=0$, is governed by

\begin{equation}
\rho(\partial_{t} +\mathbf{u}\cdot  \mathbf{\nabla} )\mathbf{u} = -\nabla p + \mathbf{\nabla}\cdot\ \mathbf{\Pi}.
\label{eq:GNS}
\end{equation}

The stress tensor here includes viscous stress $\mathbf{\Pi}_{visc} = \eta \nabla \mathbf{u}$ and active stress $   \mathbf{\Pi}_{act} = -\zeta \mathbf{Q}$, 
where $\eta$ is viscosity and $\zeta$ is activity. 
Extensile particles corresponds to a negative $\zeta$, and contractile particles corresponds to a positive $\zeta$,. 
The nematic field evolves as

\begin{equation}
    \left(\partial_{t} + \mathbf{u} \cdot \nabla \right) \mathbf{Q} = \mathbf{S}.
    \label{equation:nematic field}
\end{equation}

Here $\mathbf{S}$ is the contribution from shear flow,  which decomposes into strain and vorticity components as

\begin{equation}
    \mathbf{S} = \lambda \left(\mathbf{E}\cdot \mathbf{Q} +\mathbf{Q} \cdot \mathbf{E} \right) + \left(\mathbf{\Omega}\cdot \mathbf{Q} -\mathbf{Q} \cdot \mathbf{\Omega} \right).
\end{equation}

\begin{figure}
    \centering
    \includegraphics[width=0.7\linewidth]{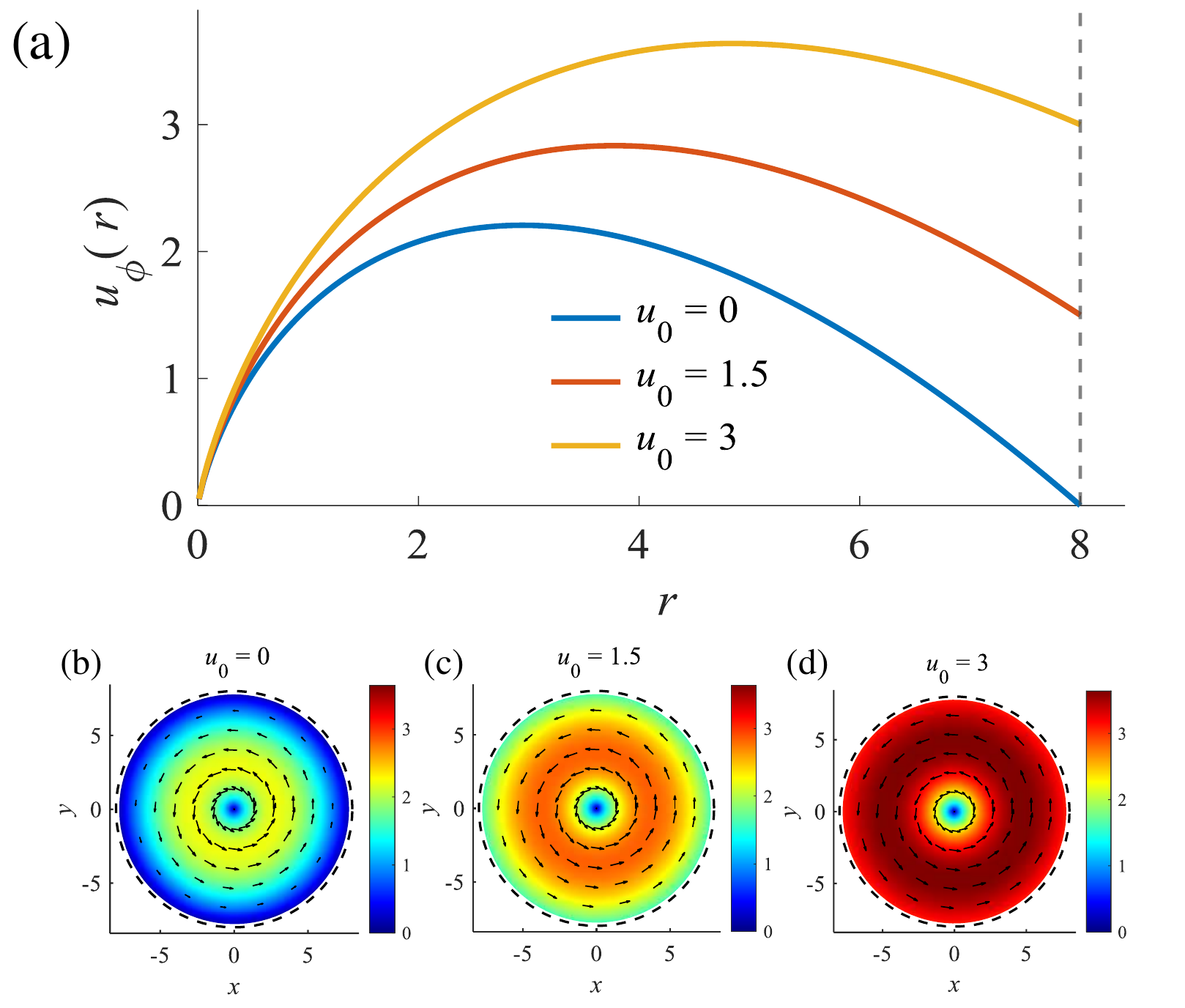}
    \caption{Velocity profile and corresponded velocity field. Fixing the confined radius $R=8$, and $\frac{\zeta \sin2\phi_{0}}{\eta}=1$}.
    \label{fig:profile and field}
\end{figure}

The flow-aligning parameter $\lambda$ controls the coupling between strain and orientation. 
The strain and vorticity are given by $\mathbf{E} = \left[\nabla \mathbf{u} + (\nabla \mathbf{u})^{T}\right]/2$ and $\mathbf{\Omega} = \left[\nabla \mathbf{u} - (\nabla \mathbf{u})^{T}\right]/2$. 
In this minimal description, we retain only the linear rotational coupling between the flow field and the orientation. 
Nonlinear coupling between the order parameter and self-generated flows is neglected by assuming a constant magnitude of the nematic order parameter.
This simplification allows us to isolate the shear-induced rotational dynamics, which dominates the stability of the vortex state.

In addition, in contrast to the full active liquid crystal framework, we neglect elastic stresses and relaxation of the orientational order induced by the molecular field.
This approximation corresponds to the flow-dominated limit of the wet active nematics, where the characteristic elastic length $l_{el}=\sqrt{K/|\zeta|}$ is much smaller than the size of the system. 
Thus, elasticity primarily regularizes defect cores without controlling the large-scale vortex dynamics~\cite{introduce_nematic_turbulence3,introduce_active_matter}.
Despite neglecting elastic relaxation, we retain the active liquid crystal description because it provides the minimal continuum framework that preserves the tensorial nature of nematic order and its symmetry-allowed coupling to flow. 

In circular confinement, nematic active particles spontaneously form a pattern with a integer defect, or such integer defects can be artificially created. 
Here, we set this integer defect as the initial state.
The direction $\mathbf{n}=(\cos{\Phi},\sin{\Phi})$ in polar coordinates, with $\Phi=\phi+\phi_{0}$, as illustrated in Fig.~\ref{fig:defect sketch}.
Thus the nematic tensor in polar coordinates has the form

\begin{equation}
    \mathbf{Q}= \frac{1}{2}\left(
    \begin{array}{cc}
    \cos2\phi_{0}  & \sin 2\phi_{0} \\
    \sin 2\phi_{0} & -\cos2\phi_{0}
    \end{array}
    \right)
\end{equation}

Active nematics capable of self-organizing into integer defects generally operate at low Reynolds number, where inertial effects are negligible.
The steady state velocity field is determined by the balance of active and viscous forces, 

\begin{equation}
    \eta\nabla^{2}\mathbf{u}=-\nabla p+\zeta\nabla\cdot\mathbf{Q}
    \label{eq:Stokes}
\end{equation}

For an axisymmetric single vortex in steady state, the incompressibility condition combined with axisymmetry $\partial_{\phi}u_{\phi}=0$, which gives $\partial_{r}(ru_{r})=0$, implying constant $u_{r}$. 
In addition, as $u_{r}$ must be finite at $r=0$, $u_{r}$must vanish everywhere. 
We thus consider consider only the tangential component $u_{\phi}$. 
Furthermore, because of the rotation symmetry, the angular coordinate $\phi$ drops out of eq.( \ref{eq:Stokes}). 
The final simplified equation is 

\begin{equation}
    \frac{\partial^{2} u_{\phi}}{\partial r^{2}}+\frac{1}{r}\frac{\partial u_{\phi}}{\partial r}-\frac{u_{\phi}}{r^{2}}=\frac{\zeta}{\eta}\frac{1}{r}\sin{2\phi_{0}}.
    \label{eq:final equation of velocity field}
\end{equation}

The natural boundary conditions at the vortex center requires the velocity to vanish, while the outer boundary is impenetrable and subject to tangential slip, with the tangential slip velocity $u_{\phi}(R)=u_{0}$. 
Under these boundary conditions, the tangential velocity profile is obtained as:

\begin{equation}
    u_{\phi}=\frac{u_{0} r}{R} + \frac{1}{2} \frac{\zeta\sin{2\phi_{0}}}{\eta}r \ln \frac{r}{R} 
    \label{eq:SteadyU_phi}
\end{equation}

A similar solution for the velocity field was obtained in controlling bacterial swarms in an artificial integer defect and active half-skyrmion. ~\cite{artificial_nematic_pattern,Half_Skyrmion}.
In active systems, inviscid-like behavior has been reported, resembling that of a superfluid between two rollers~\cite{Bacteria_Super_Fluid}. 
This motivates the application of classical hydrodynamic stability criteria to active rotating fluids. 
For physical intuition, we first draw an analogy with the classical Rayleigh criterion for rotating inviscid flows. 
Considering our confined active vortex as an active Taylor-Couette system with an infinitesimal inner cylinder, the slip velocity $u_0$ effectively sets the speed of the outer cylinder. 
The Rayleigh criterion requires that a flow be stable against axisymmetric perturbations if the angular momentum $L$ satisfies $d(L^2)/dr > 0$~\cite{Rayleigh_Criterion}. 
In the steady state of eq. (\ref{eq:SteadyU_phi}), substitute $\mathbf{L}=r \times \mathbf{u}_{\phi}$ into Rayleigh criterion, we obtain the stability condition

\begin{equation}
    \frac{4u_{0}}{R} \geq -\frac{\zeta\sin{2\phi_{0}}}{\eta}.
    \label{eq:RayleighCriterion}
\end{equation}

\begin{figure}
    \centering
    \includegraphics[width=1\linewidth]{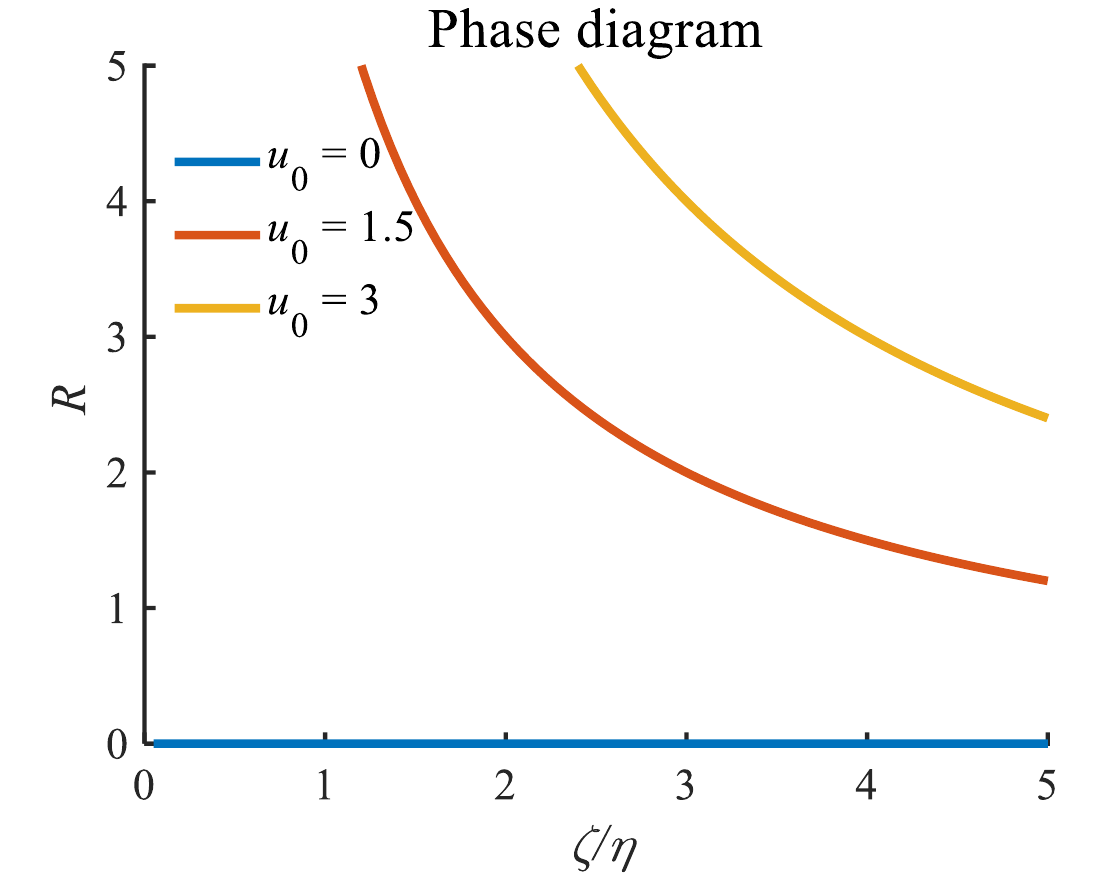}
    \caption{Stability Phase diagram. Lower part of the curve is stable region,while upper part of the curve is unstable region. As  slip velocity increases, stable region expands.}
    \label{fig:phase diagram}
\end{figure}

This result shows that a larger slip velocity $u_0$ can resist higher activity and larger size of the active system as shown in the corresponding stability phase diagram in Fig.~\ref{fig:phase diagram}.

Although the stability condition is obtained formally with an analogy to the Rayleigh stability criterion in inviscid fluids, its physical interpretation in the present low-Reynolds-number active system is different. 
A simple physical picture emerges from the coupling between the advection of the orientation field and the Jeffery-type rotation induced by the local shear flow. 
When the shear aligns along the radial direction, the Jeffery rotation tends to maintain the steady vortical alignment of the orientation. 
In contrast, if the shear reverses sign, this rotational effect changes accordingly and may locally disrupt the steady alignment. 
Increasing the slip at the boundary modifies the flow profile and suppresses such shear reversal, thereby stabilizing the system.

\section{Stability analysis}

As introduced in Section II, the Rayleigh criterion applies to centrifugal instability caused by inertia in an inviscid fluid. 
It has been reported that active systems exhibit inviscid, superfluid-like behavior at macroscopic scales~\cite{Bacteria_Super_Fluid}, reflecting the competition between viscosity and activity.
The complex flow-orientation coupling may result in entirely different instability mechanisms.
In fact, active nematics generally exist at low Reynolds number. 
In the low-Reynolds-number regime, the flow instantaneously adjusts according to the active stress through the Stokes equation, which acts as a linear constraint. 
Consequently, a perturbation in the nematic field $\delta\mathbf{Q}$ induces an instantaneous perturbation in the flow field 
$\delta\mathbf{u}$, with the two linked algebraically via the Stokes equation. 
This implies that the growth or decay of disturbances is governed solely by the dynamics of the orientational order. 
Therefore, in our linear stability analysis, we consider perturbations only to the nematic order parameter $\delta\mathbf{Q}$. 
The corresponding flow perturbation $\delta\mathbf{u}$ is understood to be slaved to $\delta\mathbf{Q}$ and does not introduce independent modes of instability.

\begin{figure*}
    \centering
    \includegraphics[width=1.02\linewidth]{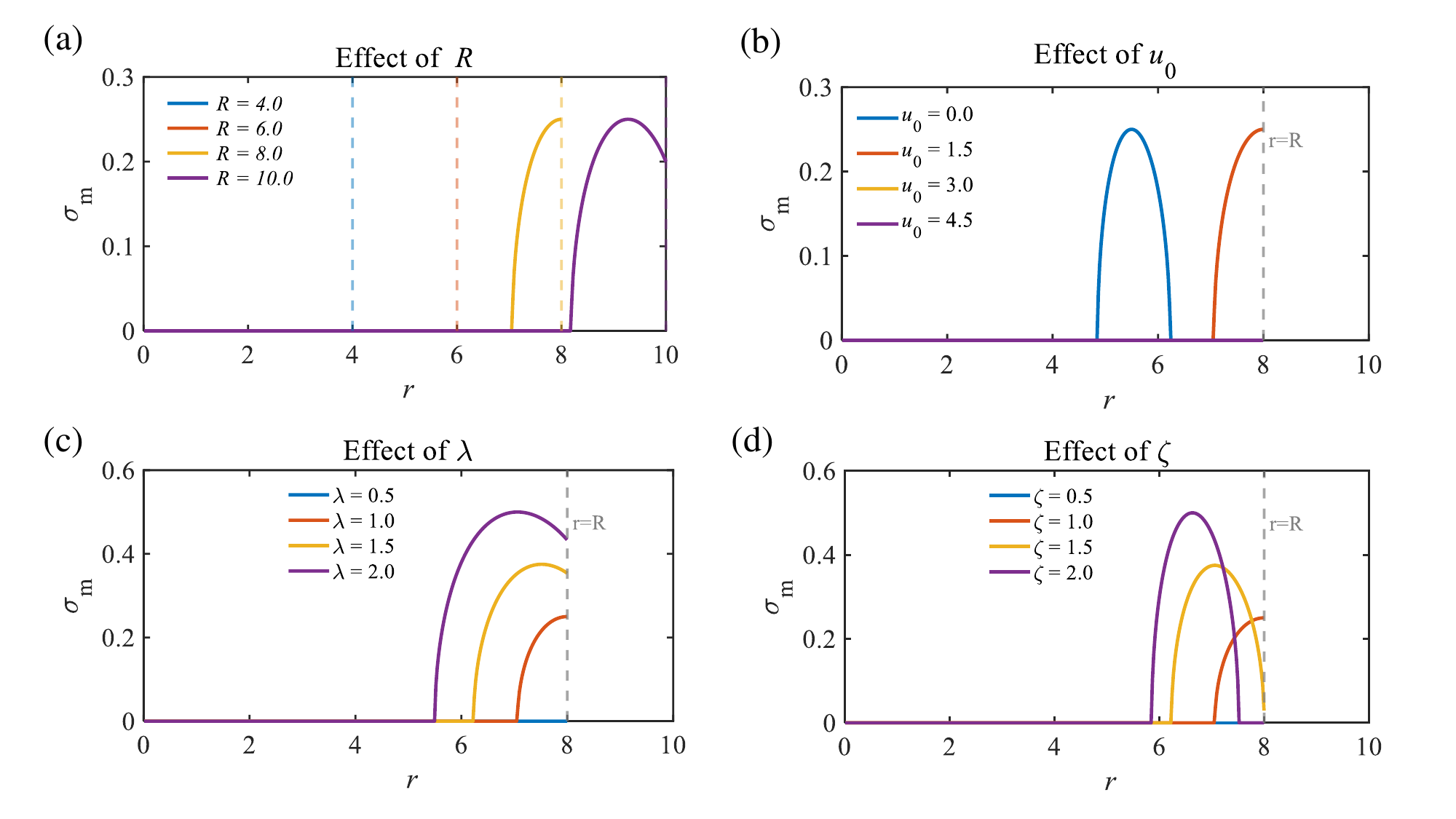}
    \caption{\textbf{Perturbation growth rate at different parameters, with fixing $\eta=1$ and $\sin2\phi_{0}=1$}. (a) shows the growth rate in different system size. The region with positive growth rate is shifted outward in radial direction with the increase of system size. Here, $\zeta=1$, $\lambda=1$ and $u_{0}=1$.
    (b) Higher slip boundary condition makes grow region move towards the edge, finally out of the confined space when the slip velocity is large enough. $\zeta=1$, $\lambda=1$ and $R=8$.
    (c) Growth rate and grow region monotonically increase as the increase of flow-orientation aligning strength. With $\zeta=1$, $u_{0}=1$ and $R=8$.
    (d) Higher activity makes higher growth rate but smaller grow region. $R=8$, $u_{0}=1$ and $\lambda=1$.}
    \label{fig4}
\end{figure*}

The small perturbation $\delta \mathbf{Q}$ is introduced into the evolution equation for the orientation field, Eq.~(\ref{equation:nematic field}). 
By retaining first-order terms in the perturbation, the linearized perturbation equation can be derived

\begin{widetext}
\begin{equation}
\left(\partial_{t}+\mathbf{u}^{0}\cdot\nabla\right)\delta\mathbf{Q}=\lambda\left(\mathbf{E}^{0}\cdot\delta\mathbf{Q}+\delta\mathbf{Q}\cdot\mathbf{E}^{0}\right)+\left(\mathbf{\Omega}^{0}\cdot\delta\mathbf{Q}-\delta\mathbf{Q}\cdot\mathbf{\Omega}^{0}\right).
\end{equation}
\end{widetext}

The perturbed nematic field is also symmetric and traceless, i.e., $\delta Q_{rr}=-\delta Q_{\phi \phi}$ and $\delta Q_{\phi r}=\delta Q_{r \phi}$.
After performing a local modal expansion of the perturbations, the evolution equations of the perturbation components become

\begin{widetext}
\begin{align}
    &\sigma_{m}\hat{Q}_{rr,m}+im\frac{u^{0}_{\phi}}{r}\hat{Q}_{rr,m}-\frac{2u^{0}_{\phi}}{r}\hat{Q}_{r\phi,m}=2\lambda E^{0}_{r\phi}\hat{Q}_{r\phi,m}+2\Omega^{0}_{r\phi}\hat{Q}_{r\phi,m}\\
    &\sigma_{m}\hat{Q}_{r\phi,m}+im\frac{u^{0}_{\phi}}{r}\hat{Q}_{r\phi,m}+\frac{2u^{0}_{\phi}}{r}\hat{Q}_{rr,m}=-2\Omega^{0}_{r\phi}\hat{Q}_{rr,m}
    \label{rp}
\end{align}
\end{widetext}

Detailed derivations and solution techniques are presented in the Appendix. 
Eq.(~\ref{eq:GrowthRate}) and Eq.(~\ref{generalized criterion}) yield a growth rate function related to ground-state flow and  flow-orientation coupling strength as

\begin{widetext}
\begin{equation}
    R(\sigma_{m})=\sqrt{\left[-\frac{4u_{0}}{R}-\frac{1}{2}\frac{\zeta \sin{2\phi_{0}}}{\eta}\left(1+4\ln \frac{r}{R}\right)\right]\left[\frac{4u_{0}}{R}+\frac{1}{2}\frac{\zeta \sin{2\phi_{0}}}{\eta}\left(1+\lambda+4\ln \frac{r}{R}\right)\right]}.
    \label{eq:GrowthRate}
\end{equation}
\end{widetext}

The perturbation growth rate $R(\sigma_m)$ depends on activity, slip velocity, flow-orientation coupling strength, and system size, as illustrated separately in Fig.~\ref{fig4}. 
The perturbation grows only beyond a certain distance from the vortex center.
This local instability arises because the underlying orientation field is spatially inhomogeneous, precluding a global Fourier mode expansion and necessitating a local angular mode decomposition instead.
In addition, the growth rate is independent of the mode $m$ in Eq. (\ref{eq:GrowthRate}), which results from the neglect of higher-order terms. 
Interestingly, the effects of activity and system size are consistent with previous studies~\cite{David_vortex_theory,artificial_nematic_pattern,vortex_instability}.
Moreover, a larger slip velocity and weaker flow-orientation coupling can suppress the occurrence of instability. 

The stabilization mechanism can be understood within a unified framework where the instability is fundamentally driven by shear and the ensuing response of the orientational field. 
All considered parameters—activity, system size, flow-orientation coupling strength, and the slip boundary condition affect stability by either modifying the shear profile or altering how closely the nematic order follows that shear. 
Specifically, the slip boundary condition reshapes the global shear distribution, which suppresses the shear gradients that would otherwise trigger orientational distortions. 
This explains why the resultant perturbation growth is spatially localized: instability only emerges where the local shear exceeds a threshold defined by the flow–orientation coupling.
This perspective also clarifies the seemingly counterintuitive observation that the instability window broadens when activity is reduced. 
A lower active stress weakens the overall shear rate, which extends the spatial region that remains stable. 

We can also derive a criterion analogous to Rayleigh's, obtained by requiring the two bracketed terms in Eq.~(\ref{eq:GrowthRate}) to have opposite signs:

\begin{equation}
    \frac{8|u_{0}|}{R} \geq \frac{|\zeta \sin{2\phi_{0}|}}{\eta} \left(\lambda+1\right).
    \label{generalized criterion}
\end{equation}

Notice that this relation does not hold when $\lambda=0$ .
When $\lambda=1$, it nicely recovers the classical inviscid criterion of Eq. (\ref{eq:RayleighCriterion}).
In the absence of elastic relaxation, the stability of the steady vortex state is controlled by the coupling between shear and orientational structure. 
The orientation remains stable only when it is locked to the shear-determined steady angle, if this alignment condition is violated, perturbations can grow.

\section{Discussion and conclusive remarks}
We have shown that the tangential slip boundary condition acts as a quantitative control parameter for the linear stability of a confined integer nematic defect accompanied by vortical flow. 
Within a hydrodynamic active liquid crystal framework, increasing slip velocity is able to substantially enlarge the outer region of the vortex where localized perturbations do not grow, in qualitative agreement with recent experiments~\cite{scale_free}. 
Four parameters, i.e. activity, system size, flow-alignment coefficient, and slip velocity, play important roles in the vortex stability through their regulation of the flow shear in its initial state or the following orientational response. 
In particular, the slip boundary condition provides a marginal correction by reshaping the azimuthal flow and suppressing shear-flow reversal,  thus shifting the stability region.
Therefore, incorporating the slip boundary condition is essential for a complete characterization of stability in confined active vortices.

By neglecting elastic relaxation in this minimal model, the analysis are restricted to the linear response of a pre-existing defect in the flow-dominated regime and does not address defect nucleation, wavelength selection, or fully nonlinear evolution. 
In this regime, orientational dynamics is primarily slaved to the local shear and vorticity, allowing the stability problem to be formulated kinematically. 
The pre-set steady state considered here is inspired by experimental observations of self-organized vortex states in circular confinement~\cite{Bacteria_vortex_RE_exp,Bacteria_vortex_RE_sim} and related artificial nematic patterns~\cite{artificial_nematic_pattern,artificial_nematic_pattern2,artificial_nematic_pattern_of_cell,mechanisim_of_artificial_nematic_pattern}.

The use of the Rayleigh criterion from classical inviscid Taylor–Couette flow serves as a qualitative analogy rather than a strict hydrodynamic limit. 
Although the appearance of slip-dependent boundaries in our phase diagram resembles trends reported in the theoretical work of Saintillan et al.~\cite{David_vortex_theory}, where analysis was performed under no-slip conditions within a self-propelled active suspension framework.
In that context, boundary backflow and the resultant steady structures arise from the balance between self-induced active stresses and viscous dissipation near rigid boundaries. 
 
Unlike the case starting from an isotropic state, our linear stability analysis shows that the perturbation growth rate displays a spatially localized distribution.
In the critical case, perturbation growth always starts near the edge.
This conclusion is qualitatively consistent with experimental observations~\cite{boundary_friction}.
The stable state we obtained is merely one with a globally zero perturbation growth rate, i.e., marginally stable. 
This is because elastic relaxation is neglected in this paper, a simplification applicable to wet active nematic systems where hydrodynamic effects dominate.
In experiments, it may be possible in the future to use a fluid-fluid interface to artificially control the slip boundary condition in order to regulate a long-lasting stable active integer defect.
For dry active matter dominated by elastic interactions (e.g., cell monolayers), the elastic term may introduce new modes, leading to the formation of, for example, fractional defects. 
In addition, it is expected to investigate in the future how boundary friction and slip velocity control dynamically dynamically control the steady-intermittent instability at critical sizes in a specific wet active system.

\begin{acknowledgments}
The authors acknowledge the funding support from the National Natural Science Foundation of China (No.12174306), the Natural Science Basic Research Program of Shaanxi (2023-JC-JQ-02), Shaanxi Academy of Fundamental Sciences (Mathematics, Physics No.23JSY024), the National Natural Science Foundation of China No. 12474197 and the Shaanxi Youth Science and Technology Star Project - 2025ZC-KJXX-51 
\end{acknowledgments}

\appendix
\section{Some details of derivation}
If the perturbation is linear, then the field can be regarded as a linear superposition of the steady state and the perturbation.
The nematic field equation of motion then becomes:
\begin{widetext}
\begin{equation}
\begin{split}
\left[\partial_{t} + \left(\mathbf{u}^{0} + \delta\mathbf{u}\right) \cdot \nabla\right]\left(\mathbf{Q}^{0} + \delta\mathbf{Q}\right)
&= \lambda\left[\left(\mathbf{E}^{0} + \delta\mathbf{E}\right) \cdot \left(\mathbf{Q}^{0} + \delta\mathbf{Q}\right) + \left(\mathbf{Q}^{0} + \delta\mathbf{Q}\right) \cdot \left(\mathbf{E}^{0} + \delta\mathbf{E}\right)^{0}\right] \\
&\quad + \left[\left(\mathbf{\Omega}^{0} + \delta\mathbf{\Omega}\right) \cdot \left(\mathbf{Q}^{0} + \delta\mathbf{Q}\right) - \left(\mathbf{Q}^{0} + \delta\mathbf{Q}\right) \cdot \left(\mathbf{\Omega}^{0} + \delta\mathbf{\Omega}\right)\right]
\end{split}
\end{equation}
\end{widetext}

By separating variables and neglecting higher-order terms $\delta \mathbf{u}\cdot\delta\mathbf{Q}$, we can get three equations:

\begin{footnotesize} 
\begin{align}
    (\partial_{t}+\mathbf{u}^{0}\cdot\nabla)\mathbf{Q}^0 
    &= \lambda(\mathbf{E}^{0}\cdot\mathbf{Q}^0+\mathbf{Q}^0\cdot\mathbf{E}^{0}) +(\mathbf{\Omega}^{0}\cdot\mathbf{Q}^0-\mathbf{Q}^0\cdot\mathbf{\Omega}^{0}) \label{eq:SteadyQ}\\
    (\partial_{t}+\delta\mathbf{u}\cdot\nabla)\mathbf{Q}^0 
    &= \lambda(\delta\mathbf{E}\cdot\mathbf{Q}^0+\mathbf{Q}^0\cdot\delta\mathbf{E}) +(\delta\mathbf{\Omega}\cdot\mathbf{Q}^0-\mathbf{Q}^0\cdot\delta\mathbf{\Omega}) \label{eq:SteadyQ0}\\
    (\partial_{t}+\mathbf{u}^{0}\cdot\nabla)\delta\mathbf{Q} 
    &= \lambda(\mathbf{E}^{0}\cdot\delta\mathbf{Q}+\delta\mathbf{Q}\cdot\mathbf{E}^{0}) +(\mathbf{\Omega}^{0}\cdot\delta\mathbf{Q}-\delta\mathbf{Q}\cdot\mathbf{\Omega}^{0}) \label{eq:SteadyQPertu}
\end{align}
\end{footnotesize}
The Eq.(\ref{eq:SteadyQ}) actually is the steady state equation. Here the shear tensor on ground state is 

\begin{align}
    & \mathbf{E}^{0}=\frac{1}{2}\left(
    \begin{array}{cc}
    0  & \partial_{r}u^{0}_{\phi}-\frac{u^{0}_{\phi}}{r} \\
    \partial_{r}u^{0}_{\phi}-\frac{u^{0}_{\phi}}{r} & 0
    \end{array}
    \right)\\
    & \mathbf{\Omega}^{0}=\frac{1}{2}\left(
    \begin{array}{cc}
    0  & \partial_{r}u^{0}_{\phi}+\frac{u^{0}_{\phi}}{r} \\
    -\partial_{r}u^{0}_{\phi}-\frac{u^{0}_{\phi}}{r} & 0
    \end{array}
    \right)
\end{align}

Note that, through Eq.(\ref{eq:SteadyQ}), we can prove that the nematic pattern in steady state does indeed exist and $\phi_{0}$ is determined by the flow-orientation alignment:

\begin{equation}
    \tan^2{\phi_{0}}=\frac{5}{2}+\frac{\lambda-4}{2\lambda+2}
\end{equation}

At low Reynolds numbers, the velocity disturbance is caused by axial disturbances and is time-independent, so  Eq. (\ref{eq:SteadyQ0}) can be neglected. 
For Eq. (\ref{eq:SteadyQPertu}), since we are not starting from an isotropic initial state, we cannot use a Fourier expansion for the disturbances.
Instead, the disturbance components are expanded as the product of radial modes and angular modes, together with an exponential time dependence, analogous to a plane-wave ansatz.
Because the initial vortex state is periodic only in the azimuthal direction, Fourier decomposition can be applied to the angular coordinate. 
In the radial direction, however, the vortex is bounded by the confinement and does not satisfy periodic boundary conditions. 
As a result, a global Fourier expansion in the radial direction is not suitable, and the perturbation analysis is carried out locally in the radial coordinate.

\begin{align}
    & \delta Q_{rr}=\sum^{\infty}_{m=0}\hat{Q}_{rr,m}\left(r\right)e^{im\phi+\sigma_{m}t}\\
    &\delta Q_{r\phi}=\sum^{\infty}_{m=0}\hat{Q}_{r\phi,m}\left(r\right)e^{im\phi+\sigma_{m}t}.
\end{align}

Due to the symmetric traceless nature of the nematic tensor, we only list two components here.
The evolution equation of these two components are:

\begin{footnotesize}
\begin{align}
    \partial_{t}\delta{Q}_{rr,m}+\frac{u^{0}_{\phi}}{r}\partial_{\phi}\delta{Q}_{rr,m}-\frac{2u^{0}_{\phi}}{r}\delta{Q}_{r\phi,m} &=2\lambda E^{0}_{r\phi}\delta{Q}_{r\phi,m}\nonumber\\
    &+2\Omega^{0}_{r\phi}\delta{Q}_{r\phi,m}
\end{align}
\end{footnotesize}

\begin{footnotesize}
\begin{equation}
    \partial_{t}\delta{Q}_{r\phi,m}+\frac{u^{0}_{\phi}}{r}\partial_{\phi}\delta{Q}_{r\phi,m}+\frac{2u^{0}_{\phi}}{r}\delta{Q}_{rr,m}=-2\Omega^{0}_{r\phi}\delta{Q}_{rr,m}
\end{equation}
\end{footnotesize}

To simplify the solution of the system of equations, we define a relative perturbation parameter $k$.

\begin{equation}
k=\frac{\hat{Q}_{r\phi,m}}{\hat{Q}_{rr,m}}=\pm \sqrt{\frac{-\frac{2u^{0}_{\phi}}{r}-\left(\partial_{r}u^{0}_{\phi}+\frac{u^{0}_{\phi}}{r}\right)}{\frac{2u^{0}_{\phi}}{r}+\lambda\left(\partial_{r}u^{0}_{\phi}-\frac{u^{0}_{\phi}}{r}\right)+\left(\partial_{r}u^{0}_{\phi}+\frac{u^{0}_{\phi}}{r}\right)}}
\end{equation}

In fact, we only take positive $k$ into account, because negative $k$ cannot trigger instability.

\begin{equation}
    \sigma_{m}=-im\frac{u^{0}_{\phi}}{r}+\frac{\left(3-\lambda\right)k}{r}u^{0}_{\phi}+\left(\lambda+1\right)k\partial_{r}u^{0}_{\phi}
\end{equation}

Substituting the steady state solution of velocity field, the Eq.~\ref{eq:GrowthRate} can be  obtained.

\nocite{*}

\bibliography{apssamp}

\end{document}